\documentclass[prl,twocolumn,showpacs]{revtex4-1}
\usepackage{graphicx}
\usepackage{amsmath}
\usepackage{amsfonts}
\usepackage{epsfig}
\usepackage{epstopdf}
\usepackage{color}
\usepackage{bm}

\def\beq{\begin{equation}}
\def\eeq{\end{equation}}

\def\beqn{\begin{eqnarray}}
\def\eeqn{\end{eqnarray}}
\def\begeqar{\begin{eqnarray}}
\def\endeqar{\end{eqnarray}}

\begin{document}

\title{Optical gyrotropy from axion electrodynamics in momentum space}
\author{Shudan Zhong}
\affiliation{Department of Physics, University of California,
Berkeley, CA 94720}
\author{Joseph Orenstein}
\affiliation{Department of Physics, University of California,
Berkeley, CA 94720}
\affiliation{Materials Sciences Division,
Lawrence Berkeley National Laboratory, Berkeley, CA 94720}
\author{Joel E. Moore}
\affiliation{Department of Physics, University of California,
Berkeley, CA 94720} \affiliation{Materials Sciences Division,
Lawrence Berkeley National Laboratory, Berkeley, CA 94720}
\date{\today}

\begin{abstract}
Several emergent phenomena and phases in solids arise from configurations of the electronic Berry phase in momentum space that are similar to gauge field configurations in real space such as magnetic monopoles.  We show that the momentum-space analogue of the ``axion electrodynamics'' term $\mathbf{E}\cdot\mathbf{B}$ plays a fundamental role in a unified theory of Berry-phase contributions to optical gyrotropy in time-reversal invariant materials and the chiral magnetic effect.  The Berry-phase mechanism predicts that the rotatory power along the optic axes of a crystal must sum to zero, a constraint beyond that stipulated by point group symmetry, but observed to high accuracy in classic experimental observations on $\alpha$-quartz.  Furthermore, the Berry mechanism provides a microscopic basis for the surface conductance at the interface between gyrotropic and nongyrotropic media.
\end{abstract}

\maketitle

The topological consequences of time-reversal symmetry breaking in two-dimensional electronic systems have been a focus of interest since the discovery of the quantum Hall effects~\cite{prangegirvin}.  Similarly interesting phenomena arise from breaking inversion symmetry (IS) in  three-dimensional systems, for example in one type of "Weyl semimetal,"~\cite{ashvinweyl,burkovbalents}, possibly realized in TaAs~\cite{taastheory1,taastheory2,taasexpt1,taasexpt2}, where IS breaking allows for non-trivial topological states that contain pairs of chiral gapless fermions.  At least in insulators, it is now widely known that there exist quantized transport phenomena as a result of topological invariance.  One goal of this paper is to demonstrate an example of topology in the optical response of {\it metals}, which can be derived using relatively simple semiclassical electron motion in low-symmetry solids.

The main effect to be discussed, natural optical activity, arises in materials that break IS but retain time-reversal symmetry.  We find two unexpected features of optical activity, one of which may already have been observed in $\alpha$-quartz, and obtain a general constraint on optical activity in some frequency ranges that impacts some designs for topological photonic devices.  Electron dynamics in such materials is subtle because, despite the lowering of spatial symmetry, the energy spectrum itself remains symmetric, \textit{i.e.}, $\epsilon({\bf k})=\epsilon(-{\bf k})$. Thus, the physics that underlies transport anomalies in such systems must involve the properties of the electronic wavefunctions themselves, rather than their energy levels.

It is by now generally understood that the wavefunction-dependent transport properties of electrons on a lattice are affected by the Berry curvature, $\bf{\Omega}(\textbf{k})$, of the Bloch states~\cite{karplusluttinger,sundaramniu,ahereview,niureview}. In the presence of a nonzero $\bf{\Omega}({\bf k})$, the semiclassical equations of motion for an electron wavepacket are modified to respect the duality between position and momentum space,
\beqn
{\bf \dot r({\bf k})} &=& {\bf v({\bf k})} + {\bf \dot k}\times{\bf \Omega({\bf k})} \cr\cr
-(\hbar/e)\bf {\dot k(\bf{r})} &=& {\bf E(\bf{r})} + {\bf \dot r} \times {\bf B(\bf{r})},
\label{eq:semiclassical}
\eeqn
where ${\bf v}({\bf k})=\hbar^{-1}\nabla_{\bf k} \epsilon({\bf k})$, ${\bf \Omega} = \nabla \times \langle u_k | -i \nabla_k | u_k \rangle$, and electron charge is $(-e)$.  From the symmetry of the modified equations of motion with position and momenta, it is clear that $\bf{\Omega}({\bf k})$ can be viewed as an effective magnetic field in momentum space.  In this paper we introduce another useful momentum-position space correspondence, involving the dual to the magnetoelectric coupling term in the Lagrangian density, that is, $\mathcal{L}_{ij}(\textbf{r},t)=\alpha_{ij} E_i B_j$.

The dual nature of the semiclassical equations suggests a corresponding tensor in momentum space,
\begin{equation}
  \mathcal{G}_{ij}(\textbf{k})=v_i(\bf k) \Omega_j(\bf k),
\end{equation}
which turns out to play a fundamental role in a unified theory of Berry-phase contributions to the transport and optical properties of inversion-breaking media.  The scalar diagonal part of $\alpha_{ij}$ is topological and referred to as ``axion electrodynamics''~\cite{wilczekaxion,qilong,essinmoorevanderbilt}, and the trace of ${\mathcal G}_{ij}$ also has a topological significance in multiple contexts.

As is clear from Eq.~\ref{eq:semiclassical} the signature of nonzero $\bf\Omega(\bf k)$ is the existence of an "anomalous" current transverse to the applied force. For ac electric fields the transverse current manifests as the phenomenon of optical gyrotropy, in which a medium exhibits a different index of refraction for left and right circularly polarized light~\cite{landau}. Gyrotropy in media with broken time-reversal symmetry is known as Faraday rotation, whereas in time-reversal symmetric systems it is usually referred to as "natural optical activity" (NOA). Below, we develop a semiclassical theory of NOA originating from the Berry curvature of inversion-breaking media, obtaining two new results.  First, the gyrotropic tensor, $g_{ij}$, that emerges from the topology of the Berry curvature is traceless.  This represents a constraint on the components of $g_{ij}$ beyond well-known relations imposed by the point-group symmetry and is therefore a signature of the Berry phase mechanism.  Tracelessness of the topological $g_{ij}$ potentially resolves an 80 year old mystery concerning NOA in $\alpha$-quartz and other materials~\cite{NOAquartz,chemla}. The second result is the existence of a surface current that flows in response to an electromagnetic wave incident at the interface between gyrotropically active and inactive media. The amplitude of this surface current is precisely that required to ensure that the rotation of the polarization of light reflected from this interface is zero, as is required by Onsager reciprocity in time-reversal invariant systems. Our result obtained from the Berry phase mechanism is the first example in which the surface current required by time-reversal symmetry emerges from a microscopically derived constitutive relation.

\begin{figure}[h]
  \centering
    \includegraphics[width=.5\textwidth]{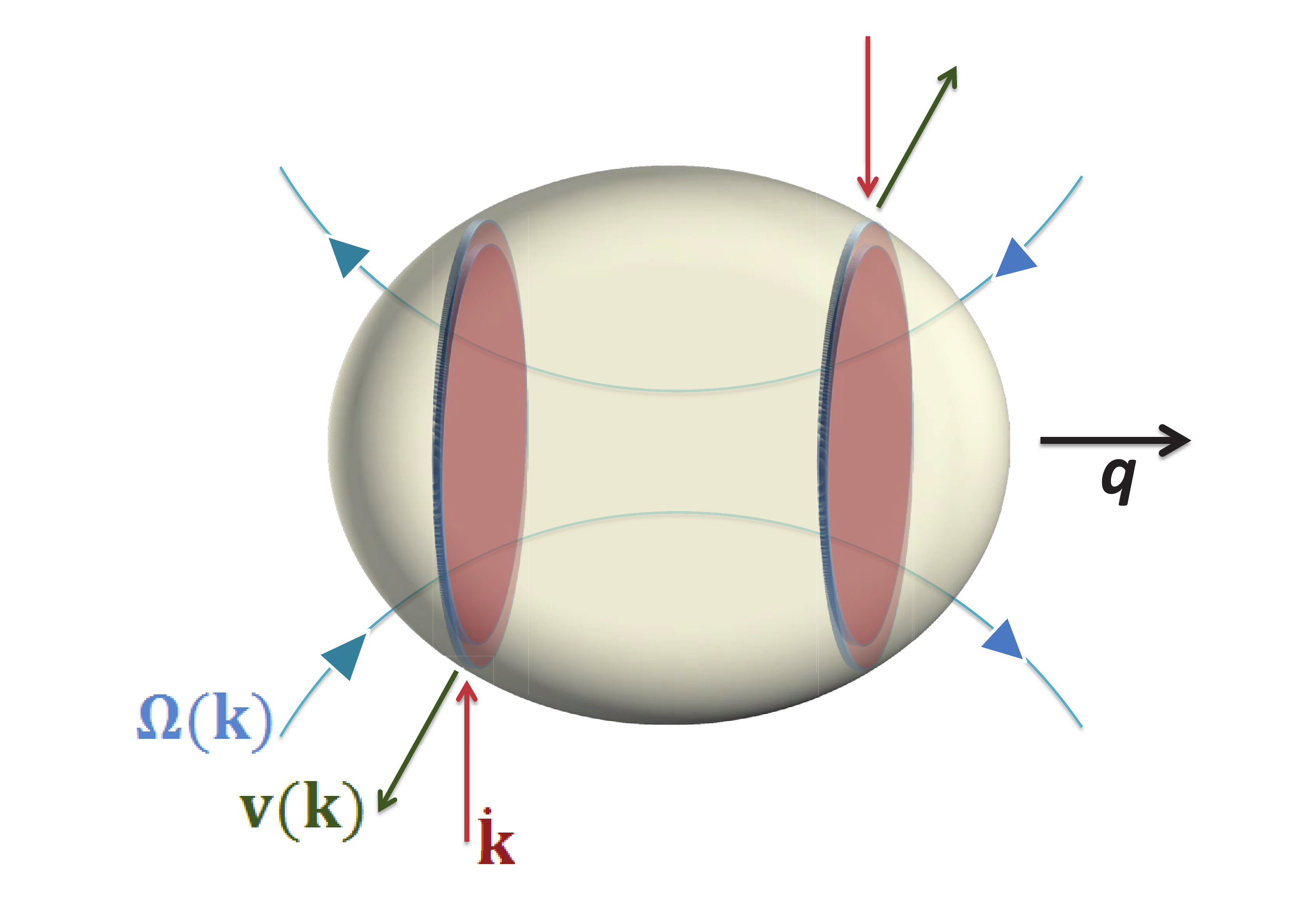}
  \caption{Momentum space representation of the Berry curvature mechanism for nonvanishing transverse current in a metal with inversion-symmetry breaking. The ellipsoid depicts a typical Fermi surface with slices oriented perpendicular to the optical wavevector.  The Berry curvature, electron velocity and acceleration to first order in optical wavevector are illustrated for two points related by time-reversal.}
\end{figure}

\textit{Tracelessness.} As a preliminary step, we rederive the Berry-phase contribution to the gyrotropy tensor in a homogeneous time-reversal-symmetric material~\cite{orensteinmoore}. We use a "Berry/Boltzmann" approach, in which the standard calculation of linear response via the Boltzmann equation is augmented by the anomalous velocity term in Eq.~\ref{eq:semiclassical}. Solving the Boltzmann equation in the relaxation time approximation yields $f^{(1)}$, the change in the distribution function to first order in the wave field $ \textbf{E}(\textbf{r},t) = \textbf{E} \exp(i\omega t - i\textbf{q}\cdot\mathbf {r}  )$,
\beq
f^{(1)} = - \nabla_k f^{(0)}\cdot\delta \bf k,
\eeq
where~\cite{dresselgruner}
\beq
\delta \textbf{k} = { e \textbf{E} \over \hbar( 1/\tau - i \omega + i\textbf{q}\cdot\bf \bf v)}.
\eeq

For simplicity, we concentrate in the following on the clean or high-frequency limit $\omega \tau \rightarrow \infty$.  The semiclassical equations are valid as long as the frequency is below that of interband transitions and neglect electron-electron interactions (although incorporating electrons at the density-functional level is simple just as for the intrinsic anomalous Hall effect).
The current that arises from the anomalous velocity is given by\footnote{We ignore a phase space volume factor $(1+{\bf B} \cdot {\Omega})^{-1}$ as ultimately it will cancel to the order required.},
\beq\label{AHcurrent}
\textbf{j} = - e \int\,{d^3k \over (2 \pi)^3}\,f^{(0)}\delta \dot{\textbf{k}}\times \bf{\Omega} ,
\eeq
where to first order in \textbf{q},
\beq\label{delta_k_dot}
\delta \dot{\textbf{k}} \approx \frac{-e\textbf{E}}{\hbar}\left(1 + \frac{\textbf{q}\cdot\bf \bf v}{\omega}\right).
\eeq
Substituting Eq. \ref{delta_k_dot} into Eq. \ref{AHcurrent} we obtain
\beq\label{constitutive}
\textbf{j} = \frac{ e^2}{\hbar }\textbf{E}\times \int\,{d^3k \over (2 \pi)^3}\,f^{(0)}\left(1+ \frac{\textbf{q}\cdot \textbf{v}}{\omega} \right)\bf \Omega.
\eeq
The \textbf{q}-independent component of the integral in Eq. \ref{constitutive} vanishes because time-reversal symmetry enforces $\bf \Omega({\bf k})=-\bf \Omega(-{\bf k})$. However, as Fig. 1 illustrates, the \textbf{q}-dependent term can be nonzero in the presence of IS breaking.  An explicit example of a tight-binding Hamiltonian with Berry curvatures of the required type was previously given~\cite{orensteinmoore}. The ellipsoid represents a typical Fermi surface and the two parallel discs are slices of momentum space perpendicular to the wavevector of the light. Focusing on two representative points related by time-reversal, we see that the acceleration $\bf \dot {k}$ to first order in \textbf{q} (shown as red arrow) is proportional to $\bf v$ and is therefore odd in \textbf{k}. Consequently the second term in the integrand of Eq. \ref{constitutive} is overall even and leads to a nonvanishing transverse current.

Next, we re-express Eq. \ref{constitutive} in the standard form for the nonlocal constitutive relation,
\begin{equation}\label{standardform}
j_i(\omega) = \sigma_{ij}(\omega) E_j + \gamma_{ijl}(\omega) \frac {dE_j}{dx_l},
\end{equation}
which relates the current to the first order of spatial derivative of the electric field ~\cite{landau}. Using Eq. \ref{constitutive},
\beq\label{gamma}
\gamma_{ijk}=\frac{-e^2}{i\hbar\omega}\int\,{d^3k \over (2 \pi)^3}\,f^{(0)} \epsilon_{ijl}\Omega_l v_k,
\eeq
where $\epsilon_{ijl}$ is the antisymmetric tensor. This response derived from the Berry curvature satisfies the condition $\gamma_{ijl}=-\gamma_{jil}$ imposed by time-reversal symmetry~\cite{landau,yudson}.

Because $\gamma_{ijk}$ is antisymmetric, the gyrotropic response is usually expressed by its dual second-rank tensor $g_{ij}$, i.e., $j_i= - i\epsilon_{ijk}g_{kl}E_j q_l$. Converting to this notation,
\beq\label{gij}
g_{ij}= - \frac{e^2}{i\hbar\omega}\int\,{d^3k \over (2 \pi)^3}\,f^{(0)}v_j\Omega_i.
\eeq
The trace of $g_{ij}$ is given by,
\beq
\sum_i g_{ii}= - \frac{e^2}{i\hbar\omega}\int\,{d^3k \over (2 \pi)^3}\,f^{(0)}\textbf{v}\cdot \bf{\Omega},
\eeq
which is zero for the ground state~\cite{niureview} or any other distribution $f^{(0)}$ depending only on energy. To see that integral over occupied states of $\bf{\Omega}\cdot \textbf{v}$ vanishes even in the presence of monopole singularities in the Berry curvature, we write, $\hbar \textbf{v}(\textbf{k})=\hat{\textbf{n}}d\varepsilon/dk_\bot$, where $\hat{\textbf{n}}$ is normal to the surface of constant energy in momentum space and $dk_\bot$ is the separation between two such surfaces whose energy differs by $d\varepsilon$. With this relation, the integral over occupied states can be written~\cite{niu_cpl},
\beq
\int\,{d^3k \over (2 \pi)^3}\,f^{(0)}\textbf{v}\cdot \mathbf{\Omega}=\int_{\varepsilon_{min}}^\mu d\varepsilon \int_\varepsilon dS\mathbf{\Omega}\cdot \hat{\textbf{n}}.
\eeq
The integral is clearly zero in the absence of singularities in $\mathbf{\Omega}$, as in this case $\nabla \cdot \mathbf{\Omega}=0$ for all $\textbf{k}$.  However, the integral still is equal to zero~\cite{niu_cpl} in the presence of singularities such as Weyl points, since
\beq
\int\,{d^3k \over (2 \pi)^3}\,f^{(0)}\textbf{v}\cdot \mathbf{\Omega}=(\mu-\varepsilon_{min})\sum_n q_n,
\eeq
which vanishes as the net monopole charge in the Brillouin zone is zero because of lattice fermion doubling~\cite{nielsenninomiya}.

Tracelessness of $g_{ij}$ has verifiable observable consequences: it is equivalent to the statement that the sum of the optical rotatory power measured along three principal axes is zero.  This rule, derived on the basis of Berry-Boltzmann physics, goes beyond the constraints imposed by point group symmetry. There are fifteen crystal classes in which non-vanishing components of $g_{ij}$ are allowed. Of these, eleven are chiral, indicating that all mirror symmetries are broken, and four have broken inversion symmetry but are not chiral.  Point group symmetry requires only these latter four classes to have traceless gyrotropic tensors.  Thus it would seem that for the other classes the observation of tracelessness would indicate the dominance of the Berry phase mechanism.

There are hints that the Berry phase mechanism is applicable to insulators as well as metals in the optical properties of $\alpha$-quartz, one of the earliest and most studied of condensed matter chiral systems~\cite{NOAquartz,chemla,wilkins}. Point group symmetry applied to $\alpha$-quartz, which belongs to crystal class 32, requires only that (in the principal axis frame) two of the three diagonal elements of $g_{ij}$ are equal, and the off-diagonal components are zero. Nevertheless, it is found experimentally that $g_{11}=g_{22}=-(1/2)g_{33}$, that is, the tensor is traceless over a broad frequency range that extends from visible to near-UV wavelengths. As it is extremely unlikely that this is accidental, there is evidence that at least in certain non-metallic systems a Berry phase-related mechanism is responsible for the gyrotropic response.  A hint is found in detailed {\it ab initio} calculations of $\alpha$-quartz and trigonal Se~\cite{wilkins}, which identify a contribution that is traceless within numerical error (the $cv$ part in that work's notation).  For a given material, measuring the trace of the gyrotropy tensor tests whether the Berry mechanism dominates other possible contributions to the gyrotropic response, for example from the Bloch electron magnetic moment (spin~\cite{tewari} or orbital).

The gyrotropic response that results from the Berry curvature is related to the question of the existence of a ``chiral magnetic effect'', a current induced by a magnetic field in the presence of pairs of Weyl nodes (related to a triangle anomaly~\cite{volovik,nielsenninomiya,ajianomaly,zyuzinburkov,sonyamamoto,parameswaranvishwanath}).  According to the semiclassical theory~\cite{niu_cpl}, the equilibrium current is
\begin{equation}\label{chiralcurrent}
  {\bf j}=\frac{e^2}{\hbar}{\bf B}\int\,{f^{(0)} d^3k \over (2 \pi)^3}{\bf \Omega} \cdot {\bf v},
\end{equation}
and is therefore zero according to the argument presented above.  However, the constitutive relation we have derived for the nonlocal current gives a closely related expression for the current that accompanies a plane electromagnetic wave. Consider a plane wave propagating along a principal axis of the crystal, which we take to be the \textbf{z}-direction. According to Eq. \ref{gij},
\begin{equation}
j_x=\frac{ -e^2}{\hbar}B_x\int\,{f^{(0)} d^3k \over (2 \pi)^3}\Omega_z v_z,
\end{equation}
where we have used the Maxwell relation $\nabla\times\mathbf{E}=-\partial \mathbf{B}/\partial t$. Thus the correct constitutive relation for the Weyl state is closely related to Eq. \ref{chiralcurrent}, but with the crucial differences that the response must be intrinsically anisotropic because of tracelessness of $\Omega_i v_j$, and the current must vanish if the magnetic field is static.

\textit{Interfacial surface current}. Combining Eq. \ref{standardform} with Maxwell's equations yields a difference in index of refraction for left and right circular polarizations, $\delta n_\pm\equiv n_+-n_-$. The latter implies a rotation of the plane of linear polarization with propagation through the medium, which is the phenomenon of NOA. At first glance, $\delta n_\pm\neq 0$ would appear to predict polarization rotation on reflection as well. The Fresnel formula for normal incidence reflection, yields a Kerr angle,
\beq
\Theta_K= \frac{\delta r_\pm}{r}=\frac{i\delta n_\pm}{n^2-1}.
\eeq
However, the expectation that $\delta r_\pm \neq 0$ has been shown to violate the general reciprocity principle for electromagnetic fields interacting with time-reversal invariant media in equilibrium~\cite{halperin}. The seeming paradox is reconciled by an historically somewhat obscure~\cite{agranovich,bokut,vinogradov} (but recently rediscovered~\cite{revisited}) constraint on the nonlocal response functions imposed by time-reversal symmetry in media in which the gyrotropic coefficient varies in space, for example at an interface. While the validity of this constraint is not in question, there has yet to be a derivation of a nonlocal constitutive relation in spatially varying chiral media that is consistent with time-reversal symmetry.

A strength of the semiclassical approach is that combining real-space and momentum-space dependence is easier than with diagrammatic methods.  We now include spatial variation of $f^{(0)}$ and show that the tracelessness of $\mathcal{G}_{ij}$ is the crucial ingredient needed to obtain a fully consistent constitutive relation at the interface.

Imagine  that either by spatial variation of chemical composition, or some form of gating (see Fig. 2), a step in potential, $V(z)$, is engineered that is sufficiently slow on the scale of the mean free path such that the semiclassical equations remain valid.  We calculate the interfacial current in response to a plane electromagnetic wave with wavevector $q_z$, to order $E_x V^\prime (z)$.

Including $V(z)$, the equilibrium distribution function $f^{(0)}$ becomes a function of $z$, leading to an extra term in the Boltzmann solution, $f^{(1)} =(\partial f^{(0)}/\partial z)\delta z$,
where,
\beq
\delta z= \frac{e}{\hbar}\frac{(\mathbf{E}\times\mathbf{\Omega)}_z}{i\omega},
\eeq
is the distance travelled in the z-direction by a carrier in one cycle of the optical frequency.  A detailed derivation of the component of $f^{(1)}$ to order $V'(z)E$ is given in the Supplemental Material. The additional current arising from the spatial variation of $f^{(0)}$ is,
\begin{equation}
\mathbf{j} ={-e^2\over i \hbar\omega}\int\,{d^3k \over (2 \pi)^3} \frac{\partial f^{(0)}}{\partial z}(\mathbf{E}\times\mathbf{\Omega)}_z \mathbf{v}(\mathbf{k}).
\end{equation}
When the width of the interface is much less that the wavelength of the light, the relevant observable is the surface sheet current,
\beq
\mathbf{K} ={- e^2\over i \hbar\omega}\int\,{d^3k \over (2 \pi)^3} \Delta f^{(0)}(\mathbf{E}\times\mathbf{\Omega)}_z \mathbf{v}(\mathbf{k}),
\label{surfacesheet}
\eeq
where $\Delta f^{(0)}$ is the change in $f^{(0)}$ across the interface. Eq. (\ref{surfacesheet}) corresponds to a constitutive relation for the interfacial current of the form, $K_i=G_{ij}E_j$, where the surface conductance is given by,
\beq
G_{ij}={ e^2\over i \hbar\omega}\int\,{d^3k \over (2 \pi)^3} \Delta f^{(0)}\epsilon_{kjz}\Omega_k(\mathbf{k})v_i(\bf k).
\eeq
If the region $z<0$ is emptied of carriers, such that we have an interface between gyrotropically active and inactive media, then $\Delta f^{(0)}=f^{(0)}$. The antisymmetric part of the surfuce conductance is
\beq
\frac{1}{2} \left( G_{xy}-G_{yx} \right ) ={e^2\over i \hbar\omega}\int\,{d^3k \over (2 \pi)^3} f^{(0)}(\Omega_x v_x+\Omega_y v_y),
\eeq
or $\frac{1}{2} \left( G_{xy}-G_{yx} \right ) = g_{zz}$, by the tracelessness of $g_{ij}$. Finally, we obtain for the antisymmetric part of the current response at the optically active/inactive interface,
\beq\label{one_half}
j_x = g_{zz}[\partial_z + \frac{1}{2}\delta(0)]E_y.
\eeq
While the factor of $1/2$ appearing in Eq. (\ref{one_half}) can be shown to be required by time-reversal symmetry~\cite{yudson}, it has not previously been derived from a microscopic or phenomenological theory. For uniaxial materials the constitutive relation Eq.~\ref{one_half} together with standard boundary conditions on the fields yields zero polarization rotation on reflection, as required by reciprocity.

\begin{figure}[h]
  \centering
    \includegraphics[width=.5\textwidth]{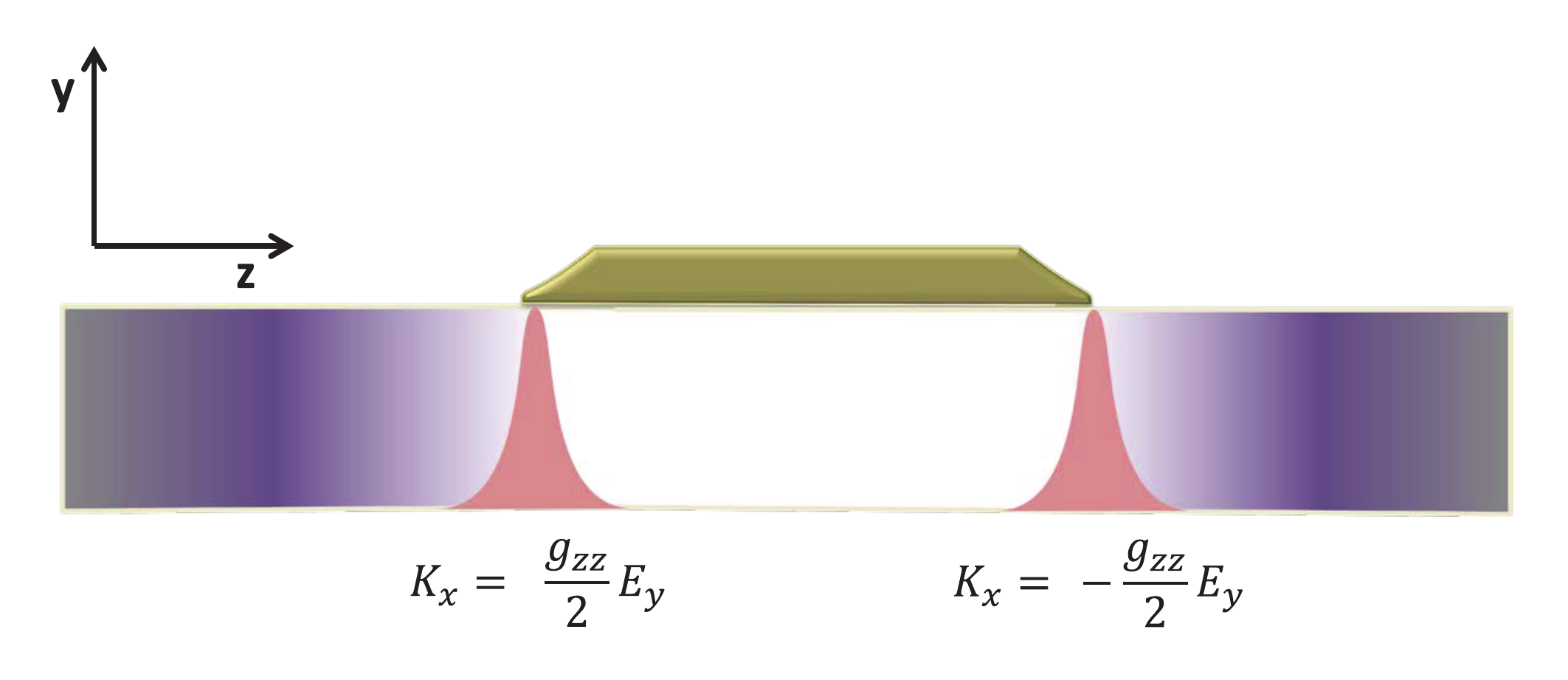}
  \caption{Illustration of a slab of an acentric metal in which electrons are depleted underneath a gate electrode. In the presence of an electromagnetic wave, counterpropagating sheet currents appear at the interfaces betweeen optically active and inactive media.}
\end{figure}

We note that in Eq.~(\ref{one_half}) time-reversal symmetry is preserved globally, but not locally. The conductivity tensor that describes the interfacial current violates Onsager reciprocity, as it is a local relation with antisymmetric off-diagonal components, \textit{i.e.}, $G_{xy}\neq G_{yx}$. Onsager reciprocity and time-reversal symmetry are restored only when considering the combined bulk and surface response.  This behavior is reminiscent of 3D topological insulators, whose surface states have an odd number of Dirac fermions, which is impossible for a 2D time-reversal symmetric system in isolation.  It would be worthwhile to understand possible additional electronic contributions to gyrotropy beyond the semiclassical static limit, as has been done for the chiral magnetic effect~\cite{changyang} in a Weyl semimetal model, where the ``uniform'' (not static) effect is nonzero but not quantized.  The Weyl semimetal TaAs~\cite{taastheory1,taastheory2,taasexpt1,taasexpt2}, along with the similar candidate materials NbAs~\cite{nbas_photo,nbas_transport,nbas_analytis} and TaP~\cite{tap_photo,tap_weyl}, breaks inversion but its space group ($l4_1md$, \#109) has point group $4mm$, which does not allow optical activity.  Either finding a different Weyl semimetal with lower symmetry, or lowering the symmetry of the TaAs family (e.g., by strain), would lead to a useful testbed for the Berry-phase contribution to gyrotropy as the magnitude of the Berry curvature is large near the Weyl points, leaving aside the possibility of open Fermi surfaces~\cite{bernevigopenweyl}.

The results presented here are important for the emerging field of topological photonics~\cite{joannopoulous}. To date, this research has focused mainly on intricately fabricated metamedia in which response functions vary periodically on the scale of the optical wavelength.  An example of a topological photonic state that can be created in this way is an analogue of the quantum Hall effect~\cite{haldaneraghu}, but interfaces between conventional materials, which require less difficult fabrication, can also support topological interface states~\cite{chiral_hyperbolic,arxiv1502.03319}.  However, in Ref.~\cite{chiral_hyperbolic} the gyrotropic response is modeled as either a pseudoscalar, or a traceful tensor with a single diagonal component, both of which are excluded by our analysis. Thus, one implication of our findings is that future analysis of chiral/nonchiral interfaces should include the traceless property of $g_{ij}$.  Finally, an important open question is the relationship between the interfacial photonic states generated by $g_{ij}$ and the time-reversal protected electron conductance, $g_{zz}$.

{\it Acknowledgements:}\,This work was primarily supported by the U.S. Department of Energy, Office of Science, Basic Energy Sciences, Materials Sciences and Engineering Division. J.E.M. acknowledges support from a Simons Investigatorship and S.Z. from NSF DMR-1206515.
\bibliography{kspaceaxion}

\begin{thebibliography}{46}%
\makeatletter
\providecommand \@ifxundefined [1]{%
 \@ifx{#1\undefined}
}%
\providecommand \@ifnum [1]{%
 \ifnum #1\expandafter \@firstoftwo
 \else \expandafter \@secondoftwo
 \fi
}%
\providecommand \@ifx [1]{%
 \ifx #1\expandafter \@firstoftwo
 \else \expandafter \@secondoftwo
 \fi
}%
\providecommand \natexlab [1]{#1}%
\providecommand \enquote  [1]{``#1''}%
\providecommand \bibnamefont  [1]{#1}%
\providecommand \bibfnamefont [1]{#1}%
\providecommand \citenamefont [1]{#1}%
\providecommand \href@noop [0]{\@secondoftwo}%
\providecommand \href [0]{\begingroup \@sanitize@url \@href}%
\providecommand \@href[1]{\@@startlink{#1}\@@href}%
\providecommand \@@href[1]{\endgroup#1\@@endlink}%
\providecommand \@sanitize@url [0]{\catcode `\\12\catcode `\$12\catcode
  `\&12\catcode `\#12\catcode `\^12\catcode `\_12\catcode `\%12\relax}%
\providecommand \@@startlink[1]{}%
\providecommand \@@endlink[0]{}%
\providecommand \url  [0]{\begingroup\@sanitize@url \@url }%
\providecommand \@url [1]{\endgroup\@href {#1}{\urlprefix }}%
\providecommand \urlprefix  [0]{URL }%
\providecommand \Eprint [0]{\href }%
\providecommand \doibase [0]{http://dx.doi.org/}%
\providecommand \selectlanguage [0]{\@gobble}%
\providecommand \bibinfo  [0]{\@secondoftwo}%
\providecommand \bibfield  [0]{\@secondoftwo}%
\providecommand \translation [1]{[#1]}%
\providecommand \BibitemOpen [0]{}%
\providecommand \bibitemStop [0]{}%
\providecommand \bibitemNoStop [0]{.\EOS\space}%
\providecommand \EOS [0]{\spacefactor3000\relax}%
\providecommand \BibitemShut  [1]{\csname bibitem#1\endcsname}%
\let\auto@bib@innerbib\@empty
\bibitem [{\citenamefont {Prange}\ and\ \citenamefont
  {Girvin}(1986)}]{prangegirvin}%
  \BibitemOpen
  \bibfield  {author} {\bibinfo {author} {\bibfnamefont {R.~E.}\ \bibnamefont
  {Prange}}\ and\ \bibinfo {author} {\bibfnamefont {S.~M.}\ \bibnamefont
  {Girvin}},\ }\href@noop {} {\emph {\bibinfo {title} {The Quantum Hall
  Effect}}}\ (\bibinfo  {publisher} {Springer},\ \bibinfo {year}
  {1986})\BibitemShut {NoStop}%
\bibitem [{\citenamefont {Wan}\ \emph {et~al.}(2011)\citenamefont {Wan},
  \citenamefont {Turner}, \citenamefont {Vishwanath},\ and\ \citenamefont
  {Savrasov}}]{ashvinweyl}%
  \BibitemOpen
  \bibfield  {author} {\bibinfo {author} {\bibfnamefont {X.}~\bibnamefont
  {Wan}}, \bibinfo {author} {\bibfnamefont {A.~M.}\ \bibnamefont {Turner}},
  \bibinfo {author} {\bibfnamefont {A.}~\bibnamefont {Vishwanath}}, \ and\
  \bibinfo {author} {\bibfnamefont {S.~Y.}\ \bibnamefont {Savrasov}},\ }\href
  {\doibase 10.1103/PhysRevB.83.205101} {\bibfield  {journal} {\bibinfo
  {journal} {Phys. Rev. B}\ }\textbf {\bibinfo {volume} {83}},\ \bibinfo
  {pages} {205101} (\bibinfo {year} {2011})}\BibitemShut {NoStop}%
\bibitem [{\citenamefont {Burkov}\ and\ \citenamefont
  {Balents}(2011)}]{burkovbalents}%
  \BibitemOpen
  \bibfield  {author} {\bibinfo {author} {\bibfnamefont {A.~A.}\ \bibnamefont
  {Burkov}}\ and\ \bibinfo {author} {\bibfnamefont {L.}~\bibnamefont
  {Balents}},\ }\href {\doibase 10.1103/PhysRevLett.107.127205} {\bibfield
  {journal} {\bibinfo  {journal} {Phys. Rev. Lett.}\ }\textbf {\bibinfo
  {volume} {107}},\ \bibinfo {pages} {127205} (\bibinfo {year}
  {2011})}\BibitemShut {NoStop}%
\bibitem [{\citenamefont {{Weng}}\ \emph {et~al.}(2015)\citenamefont {{Weng}},
  \citenamefont {{Fang}}, \citenamefont {{Fang}}, \citenamefont {{Bernevig}},\
  and\ \citenamefont {{Dai}}}]{taastheory1}%
  \BibitemOpen
  \bibfield  {author} {\bibinfo {author} {\bibfnamefont {H.}~\bibnamefont
  {{Weng}}}, \bibinfo {author} {\bibfnamefont {C.}~\bibnamefont {{Fang}}},
  \bibinfo {author} {\bibfnamefont {Z.}~\bibnamefont {{Fang}}}, \bibinfo
  {author} {\bibfnamefont {A.}~\bibnamefont {{Bernevig}}}, \ and\ \bibinfo
  {author} {\bibfnamefont {X.}~\bibnamefont {{Dai}}},\ }\href@noop {}
  {\bibfield  {journal} {\bibinfo  {journal} {Phys. Rev. X}\ }\textbf {\bibinfo
  {volume} {5}},\ \bibinfo {pages} {011029} (\bibinfo {year}
  {2015})}\BibitemShut {NoStop}%
\bibitem [{\citenamefont {{Huang}}\ \emph {et~al.}(2015)\citenamefont
  {{Huang}}, \citenamefont {{Xu}}, \citenamefont {{Belopolski}}, \citenamefont
  {{Lee}}, \citenamefont {{Chang}}, \citenamefont {{Wang}}, \citenamefont
  {{Alidoust}}, \citenamefont {{Bian}}, \citenamefont {{Neupane}},
  \citenamefont {{Bansil}}, \citenamefont {{Lin}},\ and\ \citenamefont {{Zahid
  Hasan}}}]{taastheory2}%
  \BibitemOpen
  \bibfield  {author} {\bibinfo {author} {\bibfnamefont {S.-M.}\ \bibnamefont
  {{Huang}}}, \bibinfo {author} {\bibfnamefont {S.-Y.}\ \bibnamefont {{Xu}}},
  \bibinfo {author} {\bibfnamefont {I.}~\bibnamefont {{Belopolski}}}, \bibinfo
  {author} {\bibfnamefont {C.-C.}\ \bibnamefont {{Lee}}}, \bibinfo {author}
  {\bibfnamefont {G.}~\bibnamefont {{Chang}}}, \bibinfo {author} {\bibfnamefont
  {B.}~\bibnamefont {{Wang}}}, \bibinfo {author} {\bibfnamefont
  {N.}~\bibnamefont {{Alidoust}}}, \bibinfo {author} {\bibfnamefont
  {G.}~\bibnamefont {{Bian}}}, \bibinfo {author} {\bibfnamefont
  {M.}~\bibnamefont {{Neupane}}}, \bibinfo {author} {\bibfnamefont
  {A.}~\bibnamefont {{Bansil}}}, \bibinfo {author} {\bibfnamefont
  {H.}~\bibnamefont {{Lin}}}, \ and\ \bibinfo {author} {\bibfnamefont
  {M.}~\bibnamefont {{Zahid Hasan}}},\ }\href@noop {} {\bibfield  {journal}
  {\bibinfo  {journal} {Nature Commun.}\ }\textbf {\bibinfo {volume} {6}},\
  \bibinfo {pages} {7373} (\bibinfo {year} {2015})}\BibitemShut {NoStop}%
\bibitem [{\citenamefont {Xu}\ \emph {et~al.}(2015{\natexlab{a}})\citenamefont
  {Xu}, \citenamefont {Belopolski}, \citenamefont {Alidoust}, \citenamefont
  {Neupane}, \citenamefont {Bian}, \citenamefont {Zhang}, \citenamefont
  {Sankar}, \citenamefont {Chang}, \citenamefont {Yuan}, \citenamefont {Lee},
  \citenamefont {Huang}, \citenamefont {Zheng}, \citenamefont {Ma},
  \citenamefont {Sanchez}, \citenamefont {Wang}, \citenamefont {Bansil},
  \citenamefont {Chou}, \citenamefont {Shibayev}, \citenamefont {Lin},
  \citenamefont {Jia},\ and\ \citenamefont {Hasan}}]{taasexpt1}%
  \BibitemOpen
  \bibfield  {author} {\bibinfo {author} {\bibfnamefont {S.-Y.}\ \bibnamefont
  {Xu}}, \bibinfo {author} {\bibfnamefont {I.}~\bibnamefont {Belopolski}},
  \bibinfo {author} {\bibfnamefont {N.}~\bibnamefont {Alidoust}}, \bibinfo
  {author} {\bibfnamefont {M.}~\bibnamefont {Neupane}}, \bibinfo {author}
  {\bibfnamefont {G.}~\bibnamefont {Bian}}, \bibinfo {author} {\bibfnamefont
  {C.}~\bibnamefont {Zhang}}, \bibinfo {author} {\bibfnamefont
  {R.}~\bibnamefont {Sankar}}, \bibinfo {author} {\bibfnamefont
  {G.}~\bibnamefont {Chang}}, \bibinfo {author} {\bibfnamefont
  {Z.}~\bibnamefont {Yuan}}, \bibinfo {author} {\bibfnamefont {C.-C.}\
  \bibnamefont {Lee}}, \bibinfo {author} {\bibfnamefont {S.-M.}\ \bibnamefont
  {Huang}}, \bibinfo {author} {\bibfnamefont {H.}~\bibnamefont {Zheng}},
  \bibinfo {author} {\bibfnamefont {J.}~\bibnamefont {Ma}}, \bibinfo {author}
  {\bibfnamefont {D.~S.}\ \bibnamefont {Sanchez}}, \bibinfo {author}
  {\bibfnamefont {B.}~\bibnamefont {Wang}}, \bibinfo {author} {\bibfnamefont
  {A.}~\bibnamefont {Bansil}}, \bibinfo {author} {\bibfnamefont
  {F.}~\bibnamefont {Chou}}, \bibinfo {author} {\bibfnamefont {P.~P.}\
  \bibnamefont {Shibayev}}, \bibinfo {author} {\bibfnamefont {H.}~\bibnamefont
  {Lin}}, \bibinfo {author} {\bibfnamefont {S.}~\bibnamefont {Jia}}, \ and\
  \bibinfo {author} {\bibfnamefont {M.~Z.}\ \bibnamefont {Hasan}},\ }\href@noop
  {} {\bibfield  {journal} {\bibinfo  {journal} {Science}\ } (\bibinfo {year}
  {2015}{\natexlab{a}})}\BibitemShut {NoStop}%
\bibitem [{\citenamefont {{Lv}}\ \emph {et~al.}(2015)\citenamefont {{Lv}},
  \citenamefont {{Weng}}, \citenamefont {{Fu}}, \citenamefont {{Wang}},
  \citenamefont {{Miao}}, \citenamefont {{Ma}}, \citenamefont {{Richard}},
  \citenamefont {{Huang}}, \citenamefont {{Zhao}}, \citenamefont {{Chen}},
  \citenamefont {{Fang}}, \citenamefont {{Dai}}, \citenamefont {{Qian}},\ and\
  \citenamefont {{Ding}}}]{taasexpt2}%
  \BibitemOpen
  \bibfield  {author} {\bibinfo {author} {\bibfnamefont {B.~Q.}\ \bibnamefont
  {{Lv}}}, \bibinfo {author} {\bibfnamefont {H.~M.}\ \bibnamefont {{Weng}}},
  \bibinfo {author} {\bibfnamefont {B.~B.}\ \bibnamefont {{Fu}}}, \bibinfo
  {author} {\bibfnamefont {X.~P.}\ \bibnamefont {{Wang}}}, \bibinfo {author}
  {\bibfnamefont {H.}~\bibnamefont {{Miao}}}, \bibinfo {author} {\bibfnamefont
  {J.}~\bibnamefont {{Ma}}}, \bibinfo {author} {\bibfnamefont {P.}~\bibnamefont
  {{Richard}}}, \bibinfo {author} {\bibfnamefont {X.~C.}\ \bibnamefont
  {{Huang}}}, \bibinfo {author} {\bibfnamefont {L.~X.}\ \bibnamefont {{Zhao}}},
  \bibinfo {author} {\bibfnamefont {G.~F.}\ \bibnamefont {{Chen}}}, \bibinfo
  {author} {\bibfnamefont {Z.}~\bibnamefont {{Fang}}}, \bibinfo {author}
  {\bibfnamefont {X.}~\bibnamefont {{Dai}}}, \bibinfo {author} {\bibfnamefont
  {T.}~\bibnamefont {{Qian}}}, \ and\ \bibinfo {author} {\bibfnamefont
  {H.}~\bibnamefont {{Ding}}},\ }\href@noop {} {\bibfield  {journal} {\bibinfo
  {journal} {ArXiv e-prints}\ } (\bibinfo {year} {2015})},\ \Eprint
  {http://arxiv.org/abs/1502.04684} {arXiv:1502.04684} \BibitemShut {NoStop}%
\bibitem [{\citenamefont {Karplus}\ and\ \citenamefont
  {Luttinger}(1954)}]{karplusluttinger}%
  \BibitemOpen
  \bibfield  {author} {\bibinfo {author} {\bibfnamefont {R.}~\bibnamefont
  {Karplus}}\ and\ \bibinfo {author} {\bibfnamefont {J.~M.}\ \bibnamefont
  {Luttinger}},\ }\href@noop {} {\bibfield  {journal} {\bibinfo  {journal}
  {Phys. Rev.}\ }\textbf {\bibinfo {volume} {95}},\ \bibinfo {pages} {1154}
  (\bibinfo {year} {1954})}\BibitemShut {NoStop}%
\bibitem [{\citenamefont {Sundaram}\ and\ \citenamefont
  {Niu}(1999)}]{sundaramniu}%
  \BibitemOpen
  \bibfield  {author} {\bibinfo {author} {\bibfnamefont {G.}~\bibnamefont
  {Sundaram}}\ and\ \bibinfo {author} {\bibfnamefont {Q.}~\bibnamefont {Niu}},\
  }\href@noop {} {\bibfield  {journal} {\bibinfo  {journal} {Phys. Rev. B}\
  }\textbf {\bibinfo {volume} {59}},\ \bibinfo {pages} {14915} (\bibinfo {year}
  {1999})}\BibitemShut {NoStop}%
\bibitem [{\citenamefont {Nagaosa}\ \emph {et~al.}(2010)\citenamefont
  {Nagaosa}, \citenamefont {Sinova}, \citenamefont {Onoda}, \citenamefont
  {MacDonald},\ and\ \citenamefont {Ong}}]{ahereview}%
  \BibitemOpen
  \bibfield  {author} {\bibinfo {author} {\bibfnamefont {N.}~\bibnamefont
  {Nagaosa}}, \bibinfo {author} {\bibfnamefont {J.}~\bibnamefont {Sinova}},
  \bibinfo {author} {\bibfnamefont {S.}~\bibnamefont {Onoda}}, \bibinfo
  {author} {\bibfnamefont {A.~H.}\ \bibnamefont {MacDonald}}, \ and\ \bibinfo
  {author} {\bibfnamefont {N.~P.}\ \bibnamefont {Ong}},\ }\href {\doibase
  10.1103/RevModPhys.82.1539} {\bibfield  {journal} {\bibinfo  {journal} {Rev.
  Mod. Phys.}\ }\textbf {\bibinfo {volume} {82}},\ \bibinfo {pages} {1539}
  (\bibinfo {year} {2010})}\BibitemShut {NoStop}%
\bibitem [{\citenamefont {Xiao}\ \emph {et~al.}(2010)\citenamefont {Xiao},
  \citenamefont {Chang},\ and\ \citenamefont {Niu}}]{niureview}%
  \BibitemOpen
  \bibfield  {author} {\bibinfo {author} {\bibfnamefont {D.}~\bibnamefont
  {Xiao}}, \bibinfo {author} {\bibfnamefont {M.-C.}\ \bibnamefont {Chang}}, \
  and\ \bibinfo {author} {\bibfnamefont {Q.}~\bibnamefont {Niu}},\ }\href
  {\doibase 10.1103/RevModPhys.82.1959} {\bibfield  {journal} {\bibinfo
  {journal} {Rev. Mod. Phys.}\ }\textbf {\bibinfo {volume} {82}},\ \bibinfo
  {pages} {1959} (\bibinfo {year} {2010})}\BibitemShut {NoStop}%
\bibitem [{\citenamefont {Wilczek}(1987)}]{wilczekaxion}%
  \BibitemOpen
  \bibfield  {author} {\bibinfo {author} {\bibfnamefont {F.}~\bibnamefont
  {Wilczek}},\ }\href@noop {} {\bibfield  {journal} {\bibinfo  {journal} {Phys.
  Rev. Lett.}\ }\textbf {\bibinfo {volume} {58}},\ \bibinfo {pages} {1799}
  (\bibinfo {year} {1987})}\BibitemShut {NoStop}%
\bibitem [{\citenamefont {Qi}\ \emph {et~al.}(2008)\citenamefont {Qi},
  \citenamefont {Hughes},\ and\ \citenamefont {Zhang}}]{qilong}%
  \BibitemOpen
  \bibfield  {author} {\bibinfo {author} {\bibfnamefont {X.-L.}\ \bibnamefont
  {Qi}}, \bibinfo {author} {\bibfnamefont {T.~L.}\ \bibnamefont {Hughes}}, \
  and\ \bibinfo {author} {\bibfnamefont {S.-C.}\ \bibnamefont {Zhang}},\ }\href
  {\doibase 10.1103/PhysRevB.78.195424} {\bibfield  {journal} {\bibinfo
  {journal} {Physical Review B}\ }\textbf {\bibinfo {volume} {78}},\ \bibinfo
  {eid} {195424} (\bibinfo {year} {2008})}\BibitemShut {NoStop}%
\bibitem [{\citenamefont {Essin}\ \emph {et~al.}(2009)\citenamefont {Essin},
  \citenamefont {Moore},\ and\ \citenamefont
  {Vanderbilt}}]{essinmoorevanderbilt}%
  \BibitemOpen
  \bibfield  {author} {\bibinfo {author} {\bibfnamefont {A.~M.}\ \bibnamefont
  {Essin}}, \bibinfo {author} {\bibfnamefont {J.~E.}\ \bibnamefont {Moore}}, \
  and\ \bibinfo {author} {\bibfnamefont {D.}~\bibnamefont {Vanderbilt}},\
  }\href {\doibase 10.1103/PhysRevLett.102.146805} {\bibfield  {journal}
  {\bibinfo  {journal} {Physical Review Letters}\ }\textbf {\bibinfo {volume}
  {102}},\ \bibinfo {eid} {146805} (\bibinfo {year} {2009})}\BibitemShut
  {NoStop}%
\bibitem [{\citenamefont {Landau}\ and\ \citenamefont
  {Lifshitz}(1960)}]{landau}%
  \BibitemOpen
  \bibfield  {author} {\bibinfo {author} {\bibfnamefont {L.}~\bibnamefont
  {Landau}}\ and\ \bibinfo {author} {\bibfnamefont {E.}~\bibnamefont
  {Lifshitz}},\ }\href@noop {} {\emph {\bibinfo {title} {Electrodynamics of
  Continuous Media}}}\ (\bibinfo  {publisher} {Addison-Wesley},\ \bibinfo
  {year} {1960})\BibitemShut {NoStop}%
\bibitem [{\citenamefont {Bruhat}\ and\ \citenamefont
  {Grivet}(1935)}]{NOAquartz}%
  \BibitemOpen
  \bibfield  {author} {\bibinfo {author} {\bibfnamefont {G.}~\bibnamefont
  {Bruhat}}\ and\ \bibinfo {author} {\bibfnamefont {P.}~\bibnamefont
  {Grivet}},\ }\href {\doibase 10.1103/RevModPhys.82.1539} {\bibfield
  {journal} {\bibinfo  {journal} {J. Phys.}\ }\textbf {\bibinfo {volume}
  {VI-1}},\ \bibinfo {pages} {12} (\bibinfo {year} {1935})}\BibitemShut
  {NoStop}%
\bibitem [{\citenamefont {Jerphagnon}\ and\ \citenamefont
  {Chemla}(1976)}]{chemla}%
  \BibitemOpen
  \bibfield  {author} {\bibinfo {author} {\bibfnamefont {J.}~\bibnamefont
  {Jerphagnon}}\ and\ \bibinfo {author} {\bibfnamefont {D.~S.}\ \bibnamefont
  {Chemla}},\ }\href {\doibase 10.1103/RevModPhys.82.1539} {\bibfield
  {journal} {\bibinfo  {journal} {J. Chem. Phys.}\ }\textbf {\bibinfo {volume}
  {65}},\ \bibinfo {pages} {1522} (\bibinfo {year} {1976})}\BibitemShut
  {NoStop}%
\bibitem [{\citenamefont {Orenstein}\ and\ \citenamefont
  {Moore}(2013)}]{orensteinmoore}%
  \BibitemOpen
  \bibfield  {author} {\bibinfo {author} {\bibfnamefont {J.}~\bibnamefont
  {Orenstein}}\ and\ \bibinfo {author} {\bibfnamefont {J.~E.}\ \bibnamefont
  {Moore}},\ }\href {\doibase 10.1103/PhysRevB.87.165110} {\bibfield  {journal}
  {\bibinfo  {journal} {Phys. Rev. B}\ }\textbf {\bibinfo {volume} {87}},\
  \bibinfo {pages} {165110} (\bibinfo {year} {2013})}\BibitemShut {NoStop}%
\bibitem [{\citenamefont {Dressel}\ and\ \citenamefont
  {Gruner}(2002)}]{dresselgruner}%
  \BibitemOpen
  \bibfield  {author} {\bibinfo {author} {\bibfnamefont {M.}~\bibnamefont
  {Dressel}}\ and\ \bibinfo {author} {\bibfnamefont {G.}~\bibnamefont
  {Gruner}},\ }\href@noop {} {\emph {\bibinfo {title} {Electrodynamics of
  Solids: Optical Properties of Electrons in Matter}}}\ (\bibinfo  {publisher}
  {Cambridge U. P.},\ \bibinfo {year} {2002})\BibitemShut {NoStop}%
\bibitem [{Note1()}]{Note1}%
  \BibitemOpen
  \bibinfo {note} {We ignore a phase space volume factor $(1+{\protect \bf B}
  \cdot {\Omega })^{-1}$ as ultimately it will cancel to the order
  required.}\BibitemShut {Stop}%
\bibitem [{\citenamefont {Agranovich}\ and\ \citenamefont
  {Yudson}(1973{\natexlab{a}})}]{yudson}%
  \BibitemOpen
  \bibfield  {author} {\bibinfo {author} {\bibfnamefont {V.~M.}\ \bibnamefont
  {Agranovich}}\ and\ \bibinfo {author} {\bibfnamefont {V.~I.}\ \bibnamefont
  {Yudson}},\ }\href@noop {} {\bibfield  {journal} {\bibinfo  {journal} {Optics
  Communications}\ }\textbf {\bibinfo {volume} {9}},\ \bibinfo {pages} {58}
  (\bibinfo {year} {1973}{\natexlab{a}})}\BibitemShut {NoStop}%
\bibitem [{\citenamefont {Zhou}\ \emph {et~al.}(2013)\citenamefont {Zhou},
  \citenamefont {Jiang}, \citenamefont {Niu},\ and\ \citenamefont
  {Shi}}]{niu_cpl}%
  \BibitemOpen
  \bibfield  {author} {\bibinfo {author} {\bibfnamefont {J.-H.}\ \bibnamefont
  {Zhou}}, \bibinfo {author} {\bibfnamefont {H.}~\bibnamefont {Jiang}},
  \bibinfo {author} {\bibfnamefont {Q.}~\bibnamefont {Niu}}, \ and\ \bibinfo
  {author} {\bibfnamefont {J.-R.}\ \bibnamefont {Shi}},\ }\href {\doibase
  10.1103/PhysRevB.87.165110} {\bibfield  {journal} {\bibinfo  {journal} {Chin.
  Phys. Lett.}\ }\textbf {\bibinfo {volume} {87}},\ \bibinfo {pages} {027101}
  (\bibinfo {year} {2013})}\BibitemShut {NoStop}%
\bibitem [{\citenamefont {Nielsen}\ and\ \citenamefont
  {Ninomiya}(1981)}]{nielsenninomiya}%
  \BibitemOpen
  \bibfield  {author} {\bibinfo {author} {\bibfnamefont {H.}~\bibnamefont
  {Nielsen}}\ and\ \bibinfo {author} {\bibfnamefont {M.}~\bibnamefont
  {Ninomiya}},\ }\href {\doibase 10.1103/PhysRevB.87.165110} {\bibfield
  {journal} {\bibinfo  {journal} {Nucl. Phys. B}\ }\textbf {\bibinfo {volume}
  {185}},\ \bibinfo {pages} {20} (\bibinfo {year} {1981})}\BibitemShut
  {NoStop}%
\bibitem [{\citenamefont {Zhong}\ \emph {et~al.}(1993)\citenamefont {Zhong},
  \citenamefont {Levine}, \citenamefont {Allan},\ and\ \citenamefont
  {Wilkins}}]{wilkins}%
  \BibitemOpen
  \bibfield  {author} {\bibinfo {author} {\bibfnamefont {H.}~\bibnamefont
  {Zhong}}, \bibinfo {author} {\bibfnamefont {Z.~H.}\ \bibnamefont {Levine}},
  \bibinfo {author} {\bibfnamefont {D.~C.}\ \bibnamefont {Allan}}, \ and\
  \bibinfo {author} {\bibfnamefont {J.~W.}\ \bibnamefont {Wilkins}},\ }\href
  {\doibase 10.1103/PhysRevB.48.1384} {\bibfield  {journal} {\bibinfo
  {journal} {Phys. Rev. B}\ }\textbf {\bibinfo {volume} {48}},\ \bibinfo
  {pages} {1384} (\bibinfo {year} {1993})}\BibitemShut {NoStop}%
\bibitem [{\citenamefont {Goswami}\ \emph {et~al.}(2014)\citenamefont
  {Goswami}, \citenamefont {Sharma},\ and\ \citenamefont {Tewari}}]{tewari}%
  \BibitemOpen
  \bibfield  {author} {\bibinfo {author} {\bibfnamefont {P.}~\bibnamefont
  {Goswami}}, \bibinfo {author} {\bibfnamefont {S.}~\bibnamefont {Sharma}}, \
  and\ \bibinfo {author} {\bibfnamefont {S.}~\bibnamefont {Tewari}},\
  }\href@noop {} {\bibfield  {journal} {\bibinfo  {journal} {arXiv:1404.2927}\
  } (\bibinfo {year} {2014})}\BibitemShut {NoStop}%
\bibitem [{\citenamefont {Volovik}(2003)}]{volovik}%
  \BibitemOpen
  \bibfield  {author} {\bibinfo {author} {\bibfnamefont {G.}~\bibnamefont
  {Volovik}},\ }\href@noop {} {\emph {\bibinfo {title} {The Universe in a
  Helium Droplet}}}\ (\bibinfo  {publisher} {Oxford: Clarendon Press},\
  \bibinfo {year} {2003})\BibitemShut {NoStop}%
\bibitem [{\citenamefont {Aji}(2012)}]{ajianomaly}%
  \BibitemOpen
  \bibfield  {author} {\bibinfo {author} {\bibfnamefont {V.}~\bibnamefont
  {Aji}},\ }\href {\doibase 10.1103/PhysRevB.85.241101} {\bibfield  {journal}
  {\bibinfo  {journal} {Phys. Rev. B}\ }\textbf {\bibinfo {volume} {85}},\
  \bibinfo {pages} {241101} (\bibinfo {year} {2012})}\BibitemShut {NoStop}%
\bibitem [{\citenamefont {Zyuzin}\ and\ \citenamefont
  {Burkov}(2012)}]{zyuzinburkov}%
  \BibitemOpen
  \bibfield  {author} {\bibinfo {author} {\bibfnamefont {A.~A.}\ \bibnamefont
  {Zyuzin}}\ and\ \bibinfo {author} {\bibfnamefont {A.~A.}\ \bibnamefont
  {Burkov}},\ }\href {\doibase 10.1103/PhysRevB.86.115133} {\bibfield
  {journal} {\bibinfo  {journal} {Phys. Rev. B}\ }\textbf {\bibinfo {volume}
  {86}},\ \bibinfo {pages} {115133} (\bibinfo {year} {2012})}\BibitemShut
  {NoStop}%
\bibitem [{\citenamefont {Son}\ and\ \citenamefont
  {Yamamoto}(2012)}]{sonyamamoto}%
  \BibitemOpen
  \bibfield  {author} {\bibinfo {author} {\bibfnamefont {D.~T.}\ \bibnamefont
  {Son}}\ and\ \bibinfo {author} {\bibfnamefont {N.}~\bibnamefont {Yamamoto}},\
  }\href {\doibase 10.1103/PhysRevLett.109.181602} {\bibfield  {journal}
  {\bibinfo  {journal} {Phys. Rev. Lett.}\ }\textbf {\bibinfo {volume} {109}},\
  \bibinfo {pages} {181602} (\bibinfo {year} {2012})}\BibitemShut {NoStop}%
\bibitem [{\citenamefont {Parameswaran}\ \emph {et~al.}(2014)\citenamefont
  {Parameswaran}, \citenamefont {Grover}, \citenamefont {Abanin}, \citenamefont
  {Pesin},\ and\ \citenamefont {Vishwanath}}]{parameswaranvishwanath}%
  \BibitemOpen
  \bibfield  {author} {\bibinfo {author} {\bibfnamefont {S.~A.}\ \bibnamefont
  {Parameswaran}}, \bibinfo {author} {\bibfnamefont {T.}~\bibnamefont
  {Grover}}, \bibinfo {author} {\bibfnamefont {D.~A.}\ \bibnamefont {Abanin}},
  \bibinfo {author} {\bibfnamefont {D.~A.}\ \bibnamefont {Pesin}}, \ and\
  \bibinfo {author} {\bibfnamefont {A.}~\bibnamefont {Vishwanath}},\ }\href
  {\doibase 10.1103/PhysRevX.4.031035} {\bibfield  {journal} {\bibinfo
  {journal} {Phys. Rev. X}\ }\textbf {\bibinfo {volume} {4}},\ \bibinfo {pages}
  {031035} (\bibinfo {year} {2014})}\BibitemShut {NoStop}%
\bibitem [{\citenamefont {Halperin}(1992)}]{halperin}%
  \BibitemOpen
  \bibfield  {author} {\bibinfo {author} {\bibfnamefont {B.}~\bibnamefont
  {Halperin}},\ }in\ \href@noop {} {\emph {\bibinfo {booktitle} {The Physics
  and Chemistry of Oxide Superconductors}}},\ \bibinfo {series and number}
  {Springer Verlag Proceedings of Physics Vol 60},\ \bibinfo {editor} {edited
  by\ \bibinfo {editor} {\bibfnamefont {Y.}~\bibnamefont {Iye}}\ and\ \bibinfo
  {editor} {\bibfnamefont {H.}~\bibnamefont {Yasuoka}}}\ (\bibinfo  {publisher}
  {Springer-Verlag, Berlin Heidelberg},\ \bibinfo {year} {1992})\ pp.\ \bibinfo
  {pages} {439--450}\BibitemShut {NoStop}%
\bibitem [{\citenamefont {Agranovich}\ and\ \citenamefont
  {Yudson}(1973{\natexlab{b}})}]{agranovich}%
  \BibitemOpen
  \bibfield  {author} {\bibinfo {author} {\bibfnamefont {V.}~\bibnamefont
  {Agranovich}}\ and\ \bibinfo {author} {\bibfnamefont {V.}~\bibnamefont
  {Yudson}},\ }\href {\doibase 10.1103/PhysRevB.48.1384} {\bibfield  {journal}
  {\bibinfo  {journal} {Opt. Commun.}\ }\textbf {\bibinfo {volume} {9}},\
  \bibinfo {pages} {58} (\bibinfo {year} {1973}{\natexlab{b}})}\BibitemShut
  {NoStop}%
\bibitem [{\citenamefont {Bokut}\ and\ \citenamefont
  {Serkyukov}(1974)}]{bokut}%
  \BibitemOpen
  \bibfield  {author} {\bibinfo {author} {\bibfnamefont {V.}~\bibnamefont
  {Bokut}}\ and\ \bibinfo {author} {\bibfnamefont {A.}~\bibnamefont
  {Serkyukov}},\ }\href {\doibase 10.1103/PhysRevB.48.1384} {\bibfield
  {journal} {\bibinfo  {journal} {Prikl. Spectrosk.}\ }\textbf {\bibinfo
  {volume} {20}},\ \bibinfo {pages} {677} (\bibinfo {year} {1974})}\BibitemShut
  {NoStop}%
\bibitem [{\citenamefont {Vinogradov}(2002)}]{vinogradov}%
  \BibitemOpen
  \bibfield  {author} {\bibinfo {author} {\bibfnamefont {A.}~\bibnamefont
  {Vinogradov}},\ }\href {\doibase 10.1103/PhysRevB.48.1384} {\bibfield
  {journal} {\bibinfo  {journal} {Physics Uspekhi}\ }\textbf {\bibinfo {volume}
  {45}},\ \bibinfo {pages} {331} (\bibinfo {year} {2002})}\BibitemShut
  {NoStop}%
\bibitem [{\citenamefont {Hosur}\ \emph {et~al.}(2014)\citenamefont {Hosur},
  \citenamefont {Kapitulnik}, \citenamefont {Kivelson}, \citenamefont
  {Orenstein}, \citenamefont {Raghu}, \citenamefont {Cho},\ and\ \citenamefont
  {Fried}}]{revisited}%
  \BibitemOpen
  \bibfield  {author} {\bibinfo {author} {\bibfnamefont {P.}~\bibnamefont
  {Hosur}}, \bibinfo {author} {\bibfnamefont {A.}~\bibnamefont {Kapitulnik}},
  \bibinfo {author} {\bibfnamefont {S.~A.}\ \bibnamefont {Kivelson}}, \bibinfo
  {author} {\bibfnamefont {J.}~\bibnamefont {Orenstein}}, \bibinfo {author}
  {\bibfnamefont {S.}~\bibnamefont {Raghu}}, \bibinfo {author} {\bibfnamefont
  {W.}~\bibnamefont {Cho}}, \ and\ \bibinfo {author} {\bibfnamefont
  {A.}~\bibnamefont {Fried}},\ }\href {\doibase 10.1103/PhysRevB.87.165110}
  {\bibfield  {journal} {\bibinfo  {journal} {Phys. Rev. B}\ } (\bibinfo {year}
  {2014}),\ 10.1103/PhysRevB.87.165110}\BibitemShut {NoStop}%
\bibitem [{\citenamefont {Chang}\ and\ \citenamefont {Yang}(2015)}]{changyang}%
  \BibitemOpen
  \bibfield  {author} {\bibinfo {author} {\bibfnamefont {M.-C.}\ \bibnamefont
  {Chang}}\ and\ \bibinfo {author} {\bibfnamefont {M.-F.}\ \bibnamefont
  {Yang}},\ }\href@noop {} {\bibfield  {journal} {\bibinfo  {journal} {Phys.
  Rev. B}\ }\textbf {\bibinfo {volume} {91}},\ \bibinfo {pages} {115203}
  (\bibinfo {year} {2015})}\BibitemShut {NoStop}%
\bibitem [{\citenamefont {Xu}\ \emph {et~al.}(2015{\natexlab{b}})\citenamefont
  {Xu}, \citenamefont {Alidoust}, \citenamefont {Belopolski}, \citenamefont
  {Zhang}, \citenamefont {Bian}, \citenamefont {Chang}, \citenamefont {Zheng},
  \citenamefont {Strokov}, \citenamefont {Sanchez}, \citenamefont {Chang},
  \citenamefont {Yuan}, \citenamefont {Mou}, \citenamefont {Wu}, \citenamefont
  {Huang}, \citenamefont {Lee}, \citenamefont {Huang}, \citenamefont {Wang},
  \citenamefont {Bansil}, \citenamefont {Jeng}, \citenamefont {Neupert},
  \citenamefont {Kaminski}, \citenamefont {Lin}, \citenamefont {Jia},\ and\
  \citenamefont {Hasan}}]{nbas_photo}%
  \BibitemOpen
  \bibfield  {author} {\bibinfo {author} {\bibfnamefont {S.-Y.}\ \bibnamefont
  {Xu}}, \bibinfo {author} {\bibfnamefont {N.}~\bibnamefont {Alidoust}},
  \bibinfo {author} {\bibfnamefont {I.}~\bibnamefont {Belopolski}}, \bibinfo
  {author} {\bibfnamefont {C.}~\bibnamefont {Zhang}}, \bibinfo {author}
  {\bibfnamefont {G.}~\bibnamefont {Bian}}, \bibinfo {author} {\bibfnamefont
  {T.-R.}\ \bibnamefont {Chang}}, \bibinfo {author} {\bibfnamefont
  {H.}~\bibnamefont {Zheng}}, \bibinfo {author} {\bibfnamefont
  {V.}~\bibnamefont {Strokov}}, \bibinfo {author} {\bibfnamefont {D.~S.}\
  \bibnamefont {Sanchez}}, \bibinfo {author} {\bibfnamefont {G.}~\bibnamefont
  {Chang}}, \bibinfo {author} {\bibfnamefont {Z.}~\bibnamefont {Yuan}},
  \bibinfo {author} {\bibfnamefont {D.}~\bibnamefont {Mou}}, \bibinfo {author}
  {\bibfnamefont {Y.}~\bibnamefont {Wu}}, \bibinfo {author} {\bibfnamefont
  {L.}~\bibnamefont {Huang}}, \bibinfo {author} {\bibfnamefont {C.-C.}\
  \bibnamefont {Lee}}, \bibinfo {author} {\bibfnamefont {S.-M.}\ \bibnamefont
  {Huang}}, \bibinfo {author} {\bibfnamefont {B.}~\bibnamefont {Wang}},
  \bibinfo {author} {\bibfnamefont {A.}~\bibnamefont {Bansil}}, \bibinfo
  {author} {\bibfnamefont {H.-T.}\ \bibnamefont {Jeng}}, \bibinfo {author}
  {\bibfnamefont {T.}~\bibnamefont {Neupert}}, \bibinfo {author} {\bibfnamefont
  {A.}~\bibnamefont {Kaminski}}, \bibinfo {author} {\bibfnamefont
  {H.}~\bibnamefont {Lin}}, \bibinfo {author} {\bibfnamefont {S.}~\bibnamefont
  {Jia}}, \ and\ \bibinfo {author} {\bibfnamefont {M.~Z.}\ \bibnamefont
  {Hasan}},\ }\href@noop {} {\bibfield  {journal} {\bibinfo  {journal} {ArXiv
  e-prints}\ } (\bibinfo {year} {2015}{\natexlab{b}})},\ \Eprint
  {http://arxiv.org/abs/1504.01350} {1504.01350} \BibitemShut {NoStop}%
\bibitem [{\citenamefont {Yang}\ \emph {et~al.}(2015)\citenamefont {Yang},
  \citenamefont {Li}, \citenamefont {Wang}, \citenamefont {Zhen},\ and\
  \citenamefont {an~Xu}}]{nbas_transport}%
  \BibitemOpen
  \bibfield  {author} {\bibinfo {author} {\bibfnamefont {X.}~\bibnamefont
  {Yang}}, \bibinfo {author} {\bibfnamefont {Y.}~\bibnamefont {Li}}, \bibinfo
  {author} {\bibfnamefont {Z.}~\bibnamefont {Wang}}, \bibinfo {author}
  {\bibfnamefont {Y.}~\bibnamefont {Zhen}}, \ and\ \bibinfo {author}
  {\bibfnamefont {Z.}~\bibnamefont {an~Xu}},\ }\href@noop {} {\bibfield
  {journal} {\bibinfo  {journal} {ArXiv e-prints}\ } (\bibinfo {year}
  {2015})},\ \Eprint {http://arxiv.org/abs/1506.02283} {1506.02283}
  \BibitemShut {NoStop}%
\bibitem [{\citenamefont {Moll}\ \emph {et~al.}(2015)\citenamefont {Moll},
  \citenamefont {Potter}, \citenamefont {Ramshaw}, \citenamefont {Modic},
  \citenamefont {Riggs}, \citenamefont {Zeng}, \citenamefont {Ghimire},
  \citenamefont {Bauer}, \citenamefont {Kealhofer}, \citenamefont {Nair},
  \citenamefont {Ronning},\ and\ \citenamefont {Analytis}}]{nbas_analytis}%
  \BibitemOpen
  \bibfield  {author} {\bibinfo {author} {\bibfnamefont {P.~J.}\ \bibnamefont
  {Moll}}, \bibinfo {author} {\bibfnamefont {A.~C.}\ \bibnamefont {Potter}},
  \bibinfo {author} {\bibfnamefont {B.}~\bibnamefont {Ramshaw}}, \bibinfo
  {author} {\bibfnamefont {K.}~\bibnamefont {Modic}}, \bibinfo {author}
  {\bibfnamefont {S.}~\bibnamefont {Riggs}}, \bibinfo {author} {\bibfnamefont
  {B.}~\bibnamefont {Zeng}}, \bibinfo {author} {\bibfnamefont {N.~J.}\
  \bibnamefont {Ghimire}}, \bibinfo {author} {\bibfnamefont {E.~D.}\
  \bibnamefont {Bauer}}, \bibinfo {author} {\bibfnamefont {R.}~\bibnamefont
  {Kealhofer}}, \bibinfo {author} {\bibfnamefont {N.}~\bibnamefont {Nair}},
  \bibinfo {author} {\bibfnamefont {F.}~\bibnamefont {Ronning}}, \ and\
  \bibinfo {author} {\bibfnamefont {J.~G.}\ \bibnamefont {Analytis}},\
  }\href@noop {} {\bibfield  {journal} {\bibinfo  {journal} {ArXiv e-prints}\ }
  (\bibinfo {year} {2015})},\ \Eprint {http://arxiv.org/abs/1507.06981}
  {1507.06981} \BibitemShut {NoStop}%
\bibitem [{\citenamefont {Xu}\ \emph {et~al.}(2015{\natexlab{c}})\citenamefont
  {Xu}, \citenamefont {Weng}, \citenamefont {Lv}, \citenamefont {Matt},
  \citenamefont {Park}, \citenamefont {Bisti}, \citenamefont {Strocov},
  \citenamefont {gawryluk}, \citenamefont {Pomjakushina}, \citenamefont
  {Conder}, \citenamefont {Plumb}, \citenamefont {Radovic}, \citenamefont
  {Auts}, \citenamefont {Yazyev}, \citenamefont {Fang}, \citenamefont {Dai},
  \citenamefont {Aeppli}, \citenamefont {Qian}, \citenamefont {Mesot},
  \citenamefont {Ding},\ and\ \citenamefont {Shi}}]{tap_photo}%
  \BibitemOpen
  \bibfield  {author} {\bibinfo {author} {\bibfnamefont {N.}~\bibnamefont
  {Xu}}, \bibinfo {author} {\bibfnamefont {H.~M.}\ \bibnamefont {Weng}},
  \bibinfo {author} {\bibfnamefont {B.~Q.}\ \bibnamefont {Lv}}, \bibinfo
  {author} {\bibfnamefont {C.}~\bibnamefont {Matt}}, \bibinfo {author}
  {\bibfnamefont {J.}~\bibnamefont {Park}}, \bibinfo {author} {\bibfnamefont
  {F.}~\bibnamefont {Bisti}}, \bibinfo {author} {\bibfnamefont {V.~N.}\
  \bibnamefont {Strocov}}, \bibinfo {author} {\bibfnamefont {D.}~\bibnamefont
  {gawryluk}}, \bibinfo {author} {\bibfnamefont {E.}~\bibnamefont
  {Pomjakushina}}, \bibinfo {author} {\bibfnamefont {K.}~\bibnamefont
  {Conder}}, \bibinfo {author} {\bibfnamefont {N.~C.}\ \bibnamefont {Plumb}},
  \bibinfo {author} {\bibfnamefont {M.}~\bibnamefont {Radovic}}, \bibinfo
  {author} {\bibfnamefont {G.}~\bibnamefont {Auts}}, \bibinfo {author}
  {\bibfnamefont {O.~V.}\ \bibnamefont {Yazyev}}, \bibinfo {author}
  {\bibfnamefont {Z.}~\bibnamefont {Fang}}, \bibinfo {author} {\bibfnamefont
  {X.}~\bibnamefont {Dai}}, \bibinfo {author} {\bibfnamefont {G.}~\bibnamefont
  {Aeppli}}, \bibinfo {author} {\bibfnamefont {T.}~\bibnamefont {Qian}},
  \bibinfo {author} {\bibfnamefont {J.}~\bibnamefont {Mesot}}, \bibinfo
  {author} {\bibfnamefont {H.}~\bibnamefont {Ding}}, \ and\ \bibinfo {author}
  {\bibfnamefont {M.}~\bibnamefont {Shi}},\ }\href@noop {} {\bibfield
  {journal} {\bibinfo  {journal} {ArXiv e-prints}\ } (\bibinfo {year}
  {2015}{\natexlab{c}})},\ \Eprint {http://arxiv.org/abs/1507.03983}
  {1507.03983} \BibitemShut {NoStop}%
\bibitem [{\citenamefont {Du}\ \emph {et~al.}(2015)\citenamefont {Du},
  \citenamefont {Wang}, \citenamefont {Mao}, \citenamefont {Khan},
  \citenamefont {Xu}, \citenamefont {Zhou}, \citenamefont {Zhang},
  \citenamefont {Yang}, \citenamefont {Chen}, \citenamefont {Feng},\ and\
  \citenamefont {Fang}}]{tap_weyl}%
  \BibitemOpen
  \bibfield  {author} {\bibinfo {author} {\bibfnamefont {J.}~\bibnamefont
  {Du}}, \bibinfo {author} {\bibfnamefont {H.}~\bibnamefont {Wang}}, \bibinfo
  {author} {\bibfnamefont {Q.}~\bibnamefont {Mao}}, \bibinfo {author}
  {\bibfnamefont {R.}~\bibnamefont {Khan}}, \bibinfo {author} {\bibfnamefont
  {B.}~\bibnamefont {Xu}}, \bibinfo {author} {\bibfnamefont {Y.}~\bibnamefont
  {Zhou}}, \bibinfo {author} {\bibfnamefont {Y.}~\bibnamefont {Zhang}},
  \bibinfo {author} {\bibfnamefont {J.}~\bibnamefont {Yang}}, \bibinfo {author}
  {\bibfnamefont {B.}~\bibnamefont {Chen}}, \bibinfo {author} {\bibfnamefont
  {C.}~\bibnamefont {Feng}}, \ and\ \bibinfo {author} {\bibfnamefont
  {M.}~\bibnamefont {Fang}},\ }\href@noop {} {\bibfield  {journal} {\bibinfo
  {journal} {ArXiv e-prints}\ } (\bibinfo {year} {2015})},\ \Eprint
  {http://arxiv.org/abs/1507.05246} {1507.05246} \BibitemShut {NoStop}%
\bibitem [{\citenamefont {Soluyanov}\ \emph {et~al.}(2015)\citenamefont
  {Soluyanov}, \citenamefont {Gresch}, \citenamefont {Wang}, \citenamefont
  {Wu}, \citenamefont {Troyer}, \citenamefont {Dai},\ and\ \citenamefont
  {Bernevig}}]{bernevigopenweyl}%
  \BibitemOpen
  \bibfield  {author} {\bibinfo {author} {\bibfnamefont {A.~A.}\ \bibnamefont
  {Soluyanov}}, \bibinfo {author} {\bibfnamefont {D.}~\bibnamefont {Gresch}},
  \bibinfo {author} {\bibfnamefont {Z.}~\bibnamefont {Wang}}, \bibinfo {author}
  {\bibfnamefont {Q.}~\bibnamefont {Wu}}, \bibinfo {author} {\bibfnamefont
  {M.}~\bibnamefont {Troyer}}, \bibinfo {author} {\bibfnamefont
  {X.}~\bibnamefont {Dai}}, \ and\ \bibinfo {author} {\bibfnamefont {B.~A.}\
  \bibnamefont {Bernevig}},\ }\href@noop {} {\bibfield  {journal} {\bibinfo
  {journal} {ArXiv e-prints}\ } (\bibinfo {year} {2015})},\ \Eprint
  {http://arxiv.org/abs/1507.01603} {1507.01603} \BibitemShut {NoStop}%
\bibitem [{\citenamefont {Lu}\ \emph {et~al.}(2014)\citenamefont {Lu},
  \citenamefont {Joannopoulos},\ and\ \citenamefont
  {Soljacic}}]{joannopoulous}%
  \BibitemOpen
  \bibfield  {author} {\bibinfo {author} {\bibfnamefont {L.}~\bibnamefont
  {Lu}}, \bibinfo {author} {\bibfnamefont {J.~D.}\ \bibnamefont
  {Joannopoulos}}, \ and\ \bibinfo {author} {\bibfnamefont {M.}~\bibnamefont
  {Soljacic}},\ }\href {\doibase 10.1103/PhysRevB.87.165110} {\bibfield
  {journal} {\bibinfo  {journal} {Nature Photonics}\ }\textbf {\bibinfo
  {volume} {8}},\ \bibinfo {pages} {822} (\bibinfo {year} {2014})}\BibitemShut
  {NoStop}%
\bibitem [{\citenamefont {Haldane}\ and\ \citenamefont
  {Raghu}(2008)}]{haldaneraghu}%
  \BibitemOpen
  \bibfield  {author} {\bibinfo {author} {\bibfnamefont {F.~D.~M.}\
  \bibnamefont {Haldane}}\ and\ \bibinfo {author} {\bibfnamefont
  {S.}~\bibnamefont {Raghu}},\ }\href {\doibase 10.1103/PhysRevLett.100.013904}
  {\bibfield  {journal} {\bibinfo  {journal} {Phys. Rev. Lett.}\ }\textbf
  {\bibinfo {volume} {100}},\ \bibinfo {pages} {013904} (\bibinfo {year}
  {2008})}\BibitemShut {NoStop}%
\bibitem [{\citenamefont {Gao}\ \emph {et~al.}(2015)\citenamefont {Gao},
  \citenamefont {Lawrence}, \citenamefont {Yang}, \citenamefont {Liu},
  \citenamefont {Fang}, \citenamefont {B\'eri}, \citenamefont {Li},\ and\
  \citenamefont {Zhang}}]{chiral_hyperbolic}%
  \BibitemOpen
  \bibfield  {author} {\bibinfo {author} {\bibfnamefont {W.}~\bibnamefont
  {Gao}}, \bibinfo {author} {\bibfnamefont {M.}~\bibnamefont {Lawrence}},
  \bibinfo {author} {\bibfnamefont {B.}~\bibnamefont {Yang}}, \bibinfo {author}
  {\bibfnamefont {F.}~\bibnamefont {Liu}}, \bibinfo {author} {\bibfnamefont
  {F.}~\bibnamefont {Fang}}, \bibinfo {author} {\bibfnamefont {B.}~\bibnamefont
  {B\'eri}}, \bibinfo {author} {\bibfnamefont {J.}~\bibnamefont {Li}}, \ and\
  \bibinfo {author} {\bibfnamefont {S.}~\bibnamefont {Zhang}},\ }\href
  {\doibase 10.1103/PhysRevLett.114.037402} {\bibfield  {journal} {\bibinfo
  {journal} {Phys. Rev. Lett.}\ }\textbf {\bibinfo {volume} {114}},\ \bibinfo
  {pages} {037402} (\bibinfo {year} {2015})}\BibitemShut {NoStop}%
\bibitem [{\citenamefont {Bliokh}\ \emph {et~al.}(2015)\citenamefont {Bliokh},
  \citenamefont {Smirnova},\ and\ \citenamefont {Nori}}]{arxiv1502.03319}%
  \BibitemOpen
  \bibfield  {author} {\bibinfo {author} {\bibfnamefont {K.~Y.}\ \bibnamefont
  {Bliokh}}, \bibinfo {author} {\bibfnamefont {D.}~\bibnamefont {Smirnova}}, \
  and\ \bibinfo {author} {\bibfnamefont {F.}~\bibnamefont {Nori}},\ }\href@noop
  {} {\bibfield  {journal} {\bibinfo  {journal} {Science}\ }\textbf {\bibinfo
  {volume} {348}},\ \bibinfo {pages} {1448} (\bibinfo {year}
  {2015})}\BibitemShut {NoStop}%
\end{thebibliography}%


\begin{widetext}
\begin{center}
{\bf Supplemental Material for ``Optical gyrotropy from axion electrodynamics in momentum space''}
\end{center}

For the specific example considered in the main text, a slowly varying scalar potential $V(z)$ is added to the originally uniform material. The effects of $V(z)$ come in two ways: the original distribution function will be inhomogeneous via a dependence on $z$, $f^0(z)$, and the semiclassical equations will be modified as
\begin{eqnarray}
  	\dot{\bm{r}} &=& \bm{v} - \dot{\bm{k}} \times \bm{\Omega}, \\
	\hbar \dot{\bm{k}} &=& -e\bm{E} -e\dot{\bm{r}} \times \bm{B} -V^\prime(z) \hat{z}.
\end{eqnarray}

With these we solve for $f^1$ which is the correction to $f^0$ in Boltzmann equation
\begin{equation}
\frac{\partial f^0}{\partial x_i} \dot{x}_i + \frac{\partial f^0}{\partial k_i} \dot{k}_i + \frac{\partial f^1}{\partial t} + \frac{\partial f^1}{\partial x_i} \dot{x}_i + \frac{\partial f^1}{\partial k_i} \dot{k}_i = - \frac{f^1}{\tau}.
\end{equation}

We will only consider terms linear in $\bm{E}$ field and keep up to the first order term in the spatial derivative. We also set $\tau \rightarrow \infty$. 

First, consider the electric field component in x direction $E_x = E_0 \exp(iqz - i\omega t)$. The solution
\begin{equation}
  f^1 = \frac{\partial f^0}{\partial k_x} \frac{eE_x}{\hbar(-i\omega + iqv_z)} + \frac{\partial f^0}{\partial z} \frac{eE_x \Omega_y}{i\hbar \omega} 
  \label{eqn:f1}
\end{equation}
can be seen directly by substituting it to the Boltzmann equation. We get
\begin{eqnarray}
  \frac{\partial f^1}{\partial z} \dot{z} & = & \frac{\partial^2 f^0}{\partial k_x\partial z} \frac{eE_x}{\hbar(-i\omega + iqv_z)} v_z + \frac{\partial f^0}{\partial k_x} \frac{iqeE_x}{-i\hbar\omega} v_z , \nonumber \\
  \frac{\partial f^1}{\partial k_z} \dot{k}_z & = &\frac{\partial^2 f}{\partial k_x \partial k_z} \frac{eE_x}{\hbar(-i\omega + iqv_z)} (-V^\prime(z)), \nonumber \\
	\frac{\partial f^1}{\partial t} &=& \frac{\partial f^0}{\hbar \partial k_x}eE_x + \frac{\partial f^0}{\partial k_x}\frac{iqeE_x}{i\hbar\omega}v_z - \frac{\partial f^0}{\hbar \partial z}eE_x\Omega_y ,\nonumber \\
	\frac{\partial f^0}{\partial z} \dot{z} & = & \frac{\partial f^0}{\hbar \partial z}(v_z + eE_x\Omega_y), \nonumber \\
  \frac{\partial f^0}{\partial k_x}\dot{k}_x & = &\frac{f^0}{\hbar \partial k_x} (-eE_x + ev_zB_y), \nonumber \\
	\frac{\partial f^0}{\partial k_z}\dot{k}_z & = & \frac{\partial f^0}{\hbar \partial k_z} (-ev_x B_y - V^\prime(z)).
\end{eqnarray}
The summation of these terms is zero.

Similarly, for $E_y$ and $E_z$ the solutions are
\begin{equation}
  f^1 = \frac{\partial f^0}{\partial k_y} \frac{eE_y}{\hbar( -i\omega + iqv_z) } + \frac{\partial f^0}{\partial z} \frac{-eE_y \Omega_x}{i\hbar \omega}
\end{equation}
and
\begin{equation}
  f^1 = \frac{\partial f^0}{\partial k_z} \frac{eE_z}{\hbar(-i\omega + iqv_z)}  
\end{equation}
respectively.

With these $f^1$, we compute the coefficients $\gamma$ and $\lambda$ in the standard expression of current $\vec{j}$ for optical rotation:
\begin{equation}
j_i = \epsilon_{ij} E_j + \gamma_{ijl} \nabla_l E_j + E_j \nabla_l \lambda_{ijl}.
\end{equation}
There is a minor subtlety in computing the current at finite $\tau$ (one that also appears in the semiclassical treatment of the anomalous Hall effect) that we discuss in the appendix.  Note that $\lambda_{ijl}$ can be divided into two parts: the symmetric part $\lambda^s$ and the antisymmetric part $\lambda^{an}$. The symmetric part $\nabla_l\lambda^s_{ijl}$ can be reorganized into $\epsilon_{ij}$. 

The first term in $f^1$ contributes to $\gamma$ because the steaming of electrons into or out of a certain $k$ space region will change the polarization.  The second term contributes to $\lambda$ through the normal effect of a shift of the Fermi surface. Note that, at the order we are concerned with, we should also include effects from the magnetic field but those turn out to cancel out via the tracelessness relation.

Below we list all the components explicitly.
\begin{equation}
\begin{aligned}
  \gamma_{yxz} &= {e^2 \over i \hbar\omega } \int\,{d^3k \over (2 \pi)^3} \,f_0 v_z \Omega_z,&
  \lambda_{yxz} &= {-e^2 \over i \hbar\omega } \int\,{d^3k \over (2 \pi)^3} \,f_0 v_y \Omega_y, \\
  \gamma_{xyz} &= {-e^2 \over i \hbar\omega } \int\,{d^3k \over (2 \pi)^3} \,f_0 v_z \Omega_z, &
  \lambda_{xyz} &={e^2 \over i \hbar\omega } \int\,{d^3k \over (2 \pi)^3} \,f_0 v_x \Omega_x, \\
  \gamma_{xzz} &={e^2 \over i \hbar\omega } \int\,{d^3k \over (2 \pi)^3} \,f_0 v_z \Omega_y, &
  \lambda_{xzz} &= 0, \\ 
  \gamma_{zxz} &={-e^2 \over i \hbar\omega } \int\,{d^3k \over (2 \pi)^3} \,f_0 v_z \Omega_y, & 
  \lambda_{zxz} &={-e^2 \over i \hbar\omega } \int\,{d^3k \over (2 \pi)^3} \,f_0 v_z \Omega_y, \\
  \gamma_{zyz} &={e^2 \over i \hbar\omega } \int\,{d^3k \over (2 \pi)^3} \,f_0 v_z \Omega_x, &
  \lambda_{zyz} &={e^2 \over i \hbar\omega } \int\,{d^3k \over (2 \pi)^3} \,f_0 v_z \Omega_x, \\
  \gamma_{yzz} &={-e^2 \over i \hbar\omega } \int\,{d^3k \over (2 \pi)^3} \,f_0 v_z \Omega_x, &
  \lambda_{yzz} &= 0. 
\end{aligned}
\end{equation}
We should have other components of $\lambda$ like $\lambda_{xxz}$ but these terms clearly belong to $\lambda^s$ .

The components of $\gamma$ above have the desired symmetry properties $\gamma_{ijl} = - \gamma_{jil}$ and it can also be shown that $\lambda^{an}_{ijl} = \frac{1}{2}\gamma_{ijl}$. First, by using the tracelessness relation
\begin{equation}
	\lambda^{an}_{yxz} = \frac{1}{2} (\lambda_{yxz} - \lambda_{xyz}) = {1 \over 2} {e^2 \over i \hbar\omega } \int\,{d^3k \over (2 \pi)^3} \,f_0  v_z \Omega_z
\end{equation}
is half of $\gamma_{yxz}$. The remaining relations can be seen straightforwardly, like 
\begin{equation}
	\lambda^{an}_{zxz} ={1 \over 2} {-e^2 \over i \hbar\omega } \int\,{d^3k \over (2 \pi)^3} \,f_0 v_z \Omega_y
\end{equation}
which is half of $\gamma_{zxz}$.

\end{widetext}

\end{document}